\documentclass[aps,prx,reprint,showpacs,showkeys,noeprint,longbibliography,superscriptaddress]{revtex4-1}

\usepackage{cmap}
\usepackage[utf8x]{inputenc}
\usepackage[T1]{fontenc}
\usepackage{textcomp}
\usepackage{amsmath}
\usepackage{latexsym}
\usepackage{float}
 \usepackage{amssymb}
\usepackage{graphicx}
\usepackage{hyperref}
\usepackage{subfigure}
\usepackage{xcolor}
\usepackage{url}
\usepackage{booktabs,tabularx,dcolumn}

\def\ba{\begin{eqnarray}}
\def\ea{\end{eqnarray}}
\def\be{\begin{equation}}
\def\ee{\end{equation}}
\def\bm{\begin{math}}
\def\me{\end{math}}

\def\del{\partial}

\newcommand{\dummy}

\makeatletter
\newcommand{\fmarki}{*}
\newcommand{\fmarkii}{\ensuremath{\dagger}}
\newcommand{\fmarkiii}{\ensuremath{\ddagger}}
\newcommand{\fmarkiv}{\ensuremath{\mathsection}}
\newcommand{\fmarkv}{\ensuremath{\mathparagraph}}
\newcommand{\fmarkvi}{\ensuremath{\|}}
\newcommand{\fmarkvii}{**}
\newcommand{\fmarkviii}{\ensuremath{\dagger\dagger}}
\newcommand{\fmarkix}{\ensuremath{\ddagger\ddagger}}
                
\def\@fnsymbol#1{{\ifcase#1\or \fmarki\or \fmarkii\or \fmarkiii\or \fmarkiv\or \fmarkv\or \fmarkvi\or \fmarkvii\or \fmarkviii\or \fmarkix \else\@ctrerr\fi}}
\makeatother

\begin{document}

\preprint{AIP/123-QED}

\title{Anomalous Conformations and Dynamics of Active Block Copolymers}
\author{Suman Majumder}\email[]{suman.jdv@gmail.com,  smajumder@amity.edu}
\affiliation{Amity Institute of Applied Sciences, Amity University Uttar Pradesh, Noida 201313,
India
}
\author{Subhajit Paul}\email[]{spaul@physics.du.ac.in}
\affiliation{Department of Physics and Astrophysics, University of Delhi, Delhi 110007, India
}




\date{\today}

\begin{abstract}
Heterogeneous distribution of passive and active domains in the chromosome plays a crucial role for its dynamic organization within the cell nucleus. Motivated by that here we investigate the steady-state conformation and dynamics of a model active-block copolymer using numerical simulations. Our results show that depending on the relative arrangements of the active and passive blocks, the polymer shows an unusual swelling, even larger than the corresponding fully active polymer. On the one hand, the dynamics of the full polymer show usual  enhanced diffusion and Rouse-like scaling behavior. On the other hand, individual passive and active blocks show anomalous transient super- and sub-diffusive dynamics. We characterize this anomalous dynamics in terms of the dependence of a generalized diffusion constant with the polymer length and activity strength.  
\end{abstract}


\maketitle
\section{Introduction}\label{intro}
During cell division the genomic DNA combines with proteins, viz., histone, to form a complex called chromatin. This helps in compaction of the long DNA ($\approx 1$ m)  to a relatively smaller nucleosome (typically $6-10~ \mu$m in diameter) that fits within the cell nucleus, and thereby allows the cell to divide \cite{sneppen2005,pollard2022}. The nucleosome folds to form chromatin fibres which further condense to create chromosomes, which eventually get replicated and separated during cell division. Naturally, chromatin plays a crucial role in the dynamic positioning of the chromosome within the cell nucleus. Hence, understanding the structure and dynamics of chromatin and chromosome has eluded physicists over the years \cite{van1992,sachs1995,marko1997,munkel1998,mateos2009,barbieri2012,jost2014,ganai2014,di2016,agrawal2017,shi2018,menon2020}. 

\par
Apart from the usual thermal fluctuations, it is known that athermal stochastic forces arising from local ATP-dependent energy. consumption, are inhomogeneously distributed within the chromosome \cite{weber2012}. Presence of such athermal kicks render it out of equilibrium and thus chromosome can be investigated using understandings of active matter \cite{Ramaswamy2010,Marchetti2013,elgeti2015,shaebani2020,winkler2020}. A more precise physical description can be given by  polymers comprised of monomers that are themselves active or can have activity induced by some external force. Considering the topology of many living entities, recently, a number of studies have been dedicated to the physics of active polymers \cite{kaiser2014,harder2014,kaiser2015,isele2015,isele2016,bianco2018,chaki2019,liu2019,winkler2020,das2021,paul2021,anderson2022,paul2022,paul2022activity,majumder2024}. All these studies considered a fully active polymer, i.e., all the monomers are active. However, chromosome has heterogeneous distribution of regions of inactive and active domains \cite{amitai2017,shi2018,agrawal2020}. Chromatin remodeling with this consideration translate the transcription-coupled enzymatic activity into differing levels of stochastic forces on each monomer of a polymer model of chromosome. 
\par
The conformation and dynamics of a passive polymer are characterized by specific scaling laws \cite{deGennesbook,doi1996,rubinstein2003}. For example the size of the polymer measured using the radius of gyration $R_g$ and the polymer length $N$ measured as the number of monomers, are related via the scaling $R_g\sim N^{\nu}$, where the critical exponent $\nu \approx 0.588$ in a good solvent \cite{clisby2010,clisby2016}. Similarly, the dynamics in absence of hydrodynamics, is highlighted by the Rouse scaling of the diffusion constant $D \sim N^{-1}$ \cite{rouse1953}. The Zimm scaling law $D\sim N^{-\nu}$ describes the corresponding behavior in presence of hydrodynamics \cite{zimm1956}. For active polymers too, the focus has always been on understanding these scaling laws. Theoretical and computational models of active polymers rely on introducing the activity by applying a local
force tangential to the polymer backbone \cite{bianco2018,vatin2024} or    
by choosing the monomers to be active  \cite{paul2021,paul2022,paul2022activity,majumder2024}. Depending on the kind of activity, the polymer may exhibit a coil-globule transition in a good solvent \cite{bianco2018} or a globule to coil transition in a poor solvent   \cite{paul2022activity}, which is in contrast with the behavior of a passive polymer in the respective solvents. In almost all these studies, irrespective of the model the dynamics of an active polymer is highlighted by the remarkable enhancement of the effective long-time diffusion constant $D_{\rm eff}$ \cite{bianco2018,majumder2024}. For active Brownian polymer, one also realizes a universal Rouse-like scaling, that gets hardly affected by hydrodynamic interactions \cite{majumder2024}.
\begin{figure*}[t!]
\centering
\includegraphics*[width=0.95\textwidth]{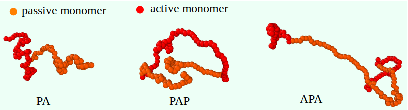}\\
\caption{\label{snapshot} Steady-state conformations of three types of active block copolymers of length $N=128$ with 1:1 ratio of active and passive monomers, obtained from simulations at $T=0.1$ and ${\rm Pe}=50$. The copolymers are named after the sequence of the active (A) and passive (P) blocks. The blocks are designed with the constraint that if there are more than one block of a particular type of monomer then each block contains the same number of monomers. Left: PA copolymer; centre: PAP copolymer; right: APA copolymer.}
\end{figure*}
\par
In this work, motivated by the partial active nature of chromosomes, using computer simulations we investigate the static and dynamic properties of active copolymers where blocks of passive monomers are connected with blocks of active monomers along the contour of the polymer (see Fig.\ \ref{snapshot}).  Our results considering three different sequences of relative arrangements of the passive and active blocks (as described in Fig.\ \ref{snapshot}) reveal unusual conformational  behavior when compared to a fully active polymer (FAP). Our novel strategy of monitoring the dynamics of the individual passive and active blocks separately reveal anomalous behavior characterized by the simultaneous existence of super-diffusion of the passive block and sub-diffusion of the active block. In spite of this diverse dynamics the long-time diffusivity of the full polymer still obeys a universal Rouse-like scaling \cite{rouse1953}. 
\par
The rest of the paper is organized in the following way. Next in Sec.\ \ref{methods}, we provide a detail description of the model used and the method of simulations. Following that in Sec.\ \ref{results} we present the results. It also includes description of the calculations of relevant observables. Finally in Sec.\ \ref{conclusion} we summarize the results and also provide an outlook to the future work.
 
\section{Model and Method}\label{methods}

We use a flexible bead-spring copolymer consisting of blocks of active and passive beads arranged according to the sequences shown in Fig.\ \ref{snapshot}. Position ${\vec{r}}_i$ of each bead follows the over-damped Langevin equation 

\begin{equation}\label{trans}
\del_t{\vec{r}}_i =\frac{ D_{\rm{tr}}}{k_BT}[f_{\rm a} \hat{n}_i-\vec{\nabla} U_i]+\sqrt{2D_{\rm{tr}}}\,\vec{\Lambda}_i^{\rm{tr}}.
\end{equation}
While for an active bead the stochastic self-propulsion force of strength $f_{\rm a}>0$ acts along the unit vector $\hat{n}_i$, using $f_{\rm a}\equiv 0$ imposes no activity for a passive bead. In Eq.\ \eqref{trans} $U_i=V_{\rm B}+V_{\rm NB}$ is the total energy, which consists of the bond energy 
\begin{equation}
V_{\rm{B}}(r_{i,i+1}) = -0.5 K(r_{i,i+1}-r_0)^2,                                                                                                                                                                                                                                                                                       \end{equation} 
with $K=100$ and the the non bonded interaction $V_{\rm NB}$, given by the standard Lennard-Jones potential 
\begin{equation}
V_{\rm{LJ}}(r_{ij}) = 4\epsilon \left[\left(\frac{\sigma}{r_{ij}}\right)^{12}- \left(\frac{\sigma}{r_{ij}}\right)^6\right],                                                                                              
\end{equation}
where $\epsilon$ is the interaction strength and $\sigma \equiv r_0\equiv1$ denotes the diameter of the monomers. For convenience during simulations, instead of the full $V_{\rm LJ}$, we use the truncated and shifted LJ potential so that the effective non-bonded interaction has the form \begin{equation}\label{nb_poten}
  V_{\rm{NB}}(r)=
\begin{cases}
  V_{\rm{LJ}}(r)-V_{\rm{LJ}}(r_c) -(r-r_c)\frac{dV_{\rm{LJ}}}{dr}\Big|_{r=r_c}  r<r_c \,,\\
0 ~~~~~~~~~ \text{otherwise}\,,
   \end{cases}
\end{equation}
where  the cut-off distance $r_c=2^{1/6}\sigma$. The orientation of the particles are updated as 
\begin{equation}
\del_t{\hat{n}}_i = \sqrt{2D_{\rm{rot}}}(\hat{n}_i\times \vec{\Lambda}_i^{\rm{rot}}).
\end{equation}
We set the ratio between the translational and rotational diffusion constants to $ {D_{\rm{tr}}}/{D_{\rm{rot}}\sigma^2} ={1}/{3}$. The vectors $\vec{\Lambda}_i^{\rm{tr}}$ and $\vec{\Lambda}_i^{\rm{rot}}$ are white Gaussian noises with zero-mean and unit-variance, and are delta-correlated over time and space. We choose the friction coefficient $\gamma \equiv 1$ and unit of time as $\tau_0=\sigma^2\gamma/\epsilon$ ($\propto 1/D_{\rm{rot}}=\Delta \sigma^2 \gamma /k_BT$ at a fixed $k_BT/\epsilon$, where $k_B$ is the Boltzmann constant).  The activity strength is expressed using the dimensionless P\'eclet number 
\begin{equation}
 {\rm Pe}=\frac{f_{\rm a}\sigma}{k_B T}.
\end{equation}
We perform simulations for different ${\rm Pe}$ at a fixed temperature $T=0.1\epsilon/k_B$, using the velocity-Verlet integration scheme with a time step of $ 10^{-4}\tau_0$ \cite{frenkel_book}. 
\begin{figure*}[t!]
\centering
\includegraphics*[width=0.95\textwidth]{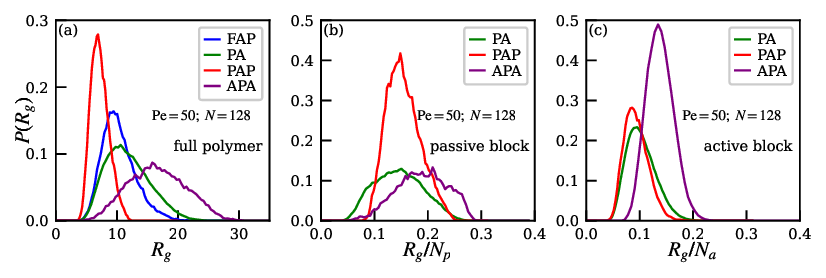}\\
\includegraphics*[width=0.95\textwidth]{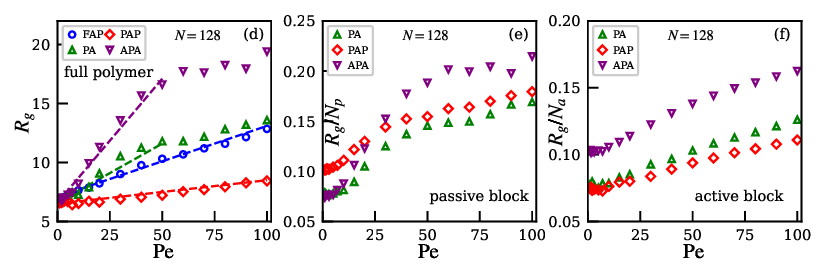}\\
\includegraphics*[width=0.95\textwidth]{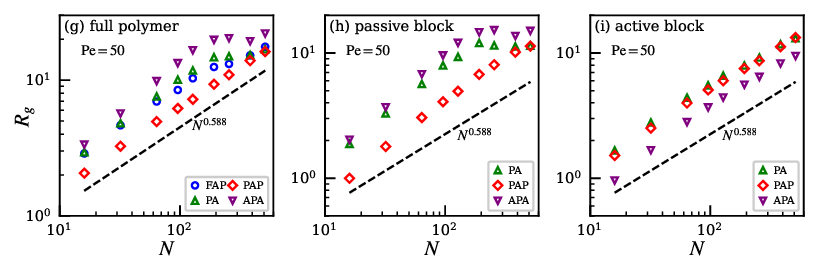}
\caption{\label{Rg_dist} Normalized distributions $P(R_g)$ of the radius of gyration of (a) the full polymer, (b) the passive block, and (c) the active block for different copolymers of fixed length $N$ at an activity ${\rm Pe}=50$. In (b) and (c) the $x$-axes are scaled by $N_p$ and $N_a$, respectively, the number of monomers present in the passive and active blocks. (d), (e), and (f) show variations of $R_g$  with ${\rm Pe}$ for $N=128$.  Plots in (g),(h), and (i) show the variation of $R_g$ with $N$ at ${\rm Pe}=50$. 
The dashed lines in (d) are linear fits to the data. The dashed line in each of (g), (h), and (i) represents the scaling of a passive polymer in a good solvent.}
\end{figure*}
\begin{figure*}[t!]
\centering
\includegraphics*[width=0.95\textwidth]{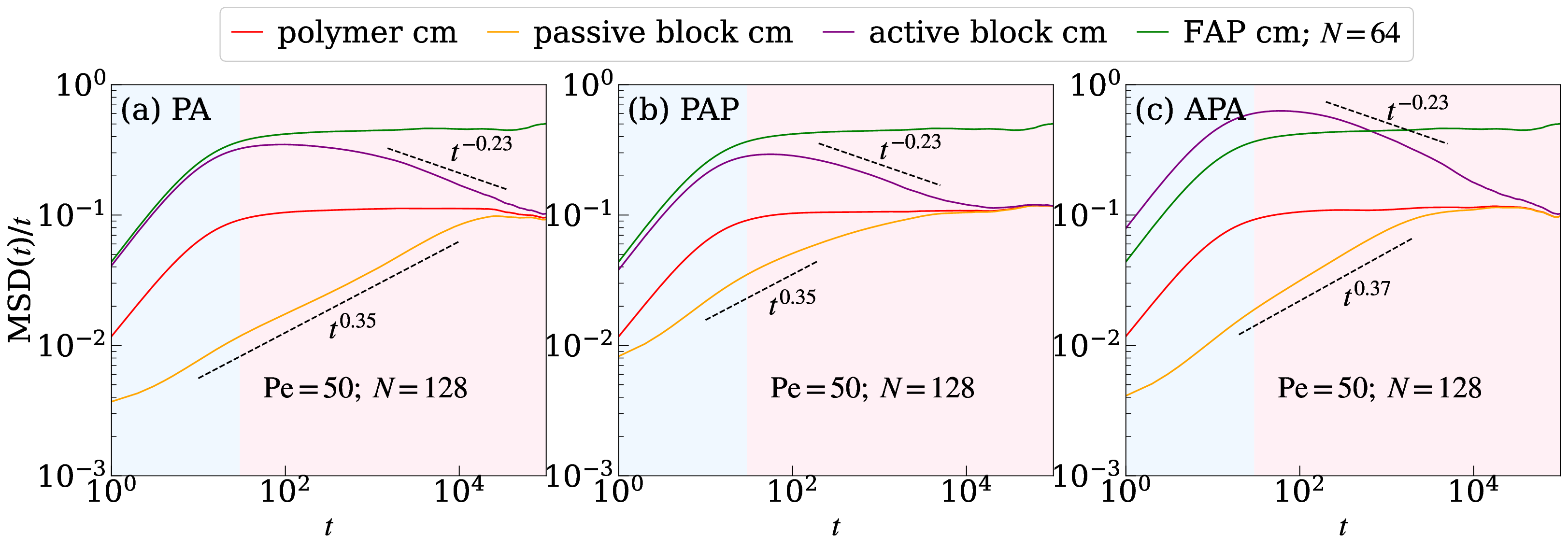}\\
\caption{\label{MSD} ${\rm MSD}(t)/t$ for the cm of the full polymer, the passive block, and the active block for different copolymers of length $N=128$  at ${\rm Pe=50}$, on a double-log scale. For a comparison, data for the cm of an FAP (green solid line) with $N=64$ is also included. The dashed lines are different power laws indicating the transient super- and sub-diffusive behavior. The shades distinguish the late-time behavior from the initial ballistic behavior of the cm of the FAP.}
\end{figure*}
\section{Results}\label{results}
We start with the conformations of active block copolymers. Fig.\ \ref{snapshot} shows typical steady-state conformations for the three types of copolymers of length $N=128$ at ${\rm Pe}=50$. As expected, with increasing ${\rm Pe}$ all the copolymers get extended. On comparing with a fully active polymer (FAP) at the same ${\rm Pe}$, surprisingly  copolymers APA and PA look more extended, while PAP appears to be shorter.  To confirm this observation we measure the size of the polymer in terms of the radius of gyration 
\begin{equation}
 R_g=\left \langle \sqrt{\frac{1}{2N^2}\sum_{i,j}(\vec{r}_i-\vec{r}_j)^2}\right \rangle, 
\end{equation}
where $\langle \dots \rangle$ denotes averaging over steady state and independent simulation runs. The distributions $P(R_g)$ at ${\rm Pe}=50$ for $N=128$ are presented in Fig.\ \ref{Rg_dist}(a). While for PAP the peak is at a smaller value ($R_g \approx 7$) than for FAP ($R_g \approx 10$), for both PA and APA the corresponding peaks are at larger $R_g$, confirming their anomalously larger swelling than a FAP. For APA, the peak is almost at a value ($R_g \approx 18$) twice than that for FAP. The variation of $R_g$ with ${\rm Pe}$ presented in Fig.\ \ref{Rg_dist}(d) reveals a more concrete picture. For ${\rm Pe}< 10$, $R_g$ for different copolymers show marginal differences. Once the activity is of considerable strength, i.e., for ${\rm Pe}>10$ differences between them show up. FAP and PAP show linear increase in $R_g$ with ${\rm Pe}$ for the full range, as illustrated by the corresponding fitted dashed lines. The data for PA show an initial steeper but linear increase until ${\rm Pe}=50$, following which it almost catch up with the data for FAP. Interestingly, the data for APA show an even steeper linear increase until ${\rm Pe}=50$, beyond which it almost saturates.

\par
The unexpectedly larger $R_g$ of the APA copolymer than a FAP can be phenomenologically understood from a careful observation of the steady-state trajectory. In an APA copolymer the two active blocks pull the passive block from both sides, resulting in forces that propagate via the bonds connecting a passive monomer with an active monomer on either end of the passive block. The thermal fluctuations of the passive monomers oppose the pulling forces, however, are much weaker comparatively at $T=0.1$. For the FAP case, the active forces of the  monomers away from the ends can easily cancel out the pulling force generated by the end monomers, and thus it behaves like a self-avoiding polymer. For the PA copolymer the pulling occurs from one end only, hence, the extension of the passive block is not as much resulting in a size comparable to the FAP. To consolidate this speculation we calculate $R_g$ of individual active and passive blocks. Note that if there are two blocks of the same type of monomer then the the presented $R_g$ is an average of the two blocks. The corresponding distributions are presented in Figs.\ \ref{Rg_dist}(b) and (c). There the $x$-axes are scaled respectively by the number of monomers present in each of the blocks. Indeed, for all the copolymers the swellings for the passive blocks are greater than the active blocks. Among them, APA copolymer show maximum swelling for both active and passive blocks, whereas data for PA and PAP copolymers are comparable. The corresponding variations with ${\rm Pe}$ presented in Figs.\ \ref{Rg_dist}(e) and (f) also reveal the same fact of larger swelling for the passive blocks than the active ones. Interestingly, the variations are quite non-monotonic with increasing ${\rm Pe}$.  At smaller ${\rm Pe}$, PA and APA have comparable sizes before they start deviating from each other. For larger ${\rm Pe}$, however, PA and PAP have comparable sizes. In case of APA and PA, the similarity in the behavior of $R_g$ for the full polymer and the passive block with increasing ${\rm Pe}$ suggests that indeed the overall size of the polymer is mostly guided by the behavior of the passive block. Likewise, for PAP the full polymer behavior is similar to the behavior of the active block. Same set of similarities can also be spotted for the $N$ dependence of $R_g$, as depicted in Figs.\ \ref{Rg_dist}(g)-(i). Except for the APA, data for $R_g$ of the full polymer [Fig.\ \ref{Rg_dist}(g)] more or less obey $R_g\sim N^{0.588}$ scaling. The same is obeyed by the $R_g$ of the active blocks [Fig.\ \ref{Rg_dist}(h)] for all the copolymers. However, for the passive blocks in all cases the data show significant deviation from self-avoiding scaling behavior [see Fig.\ \ref{Rg_dist}(i)]. Again for APA, the behavior of the passive block is similar to what is observed for the full polymer. 
\begin{figure*}[t!]
\centering
\includegraphics*[width=0.95\textwidth]{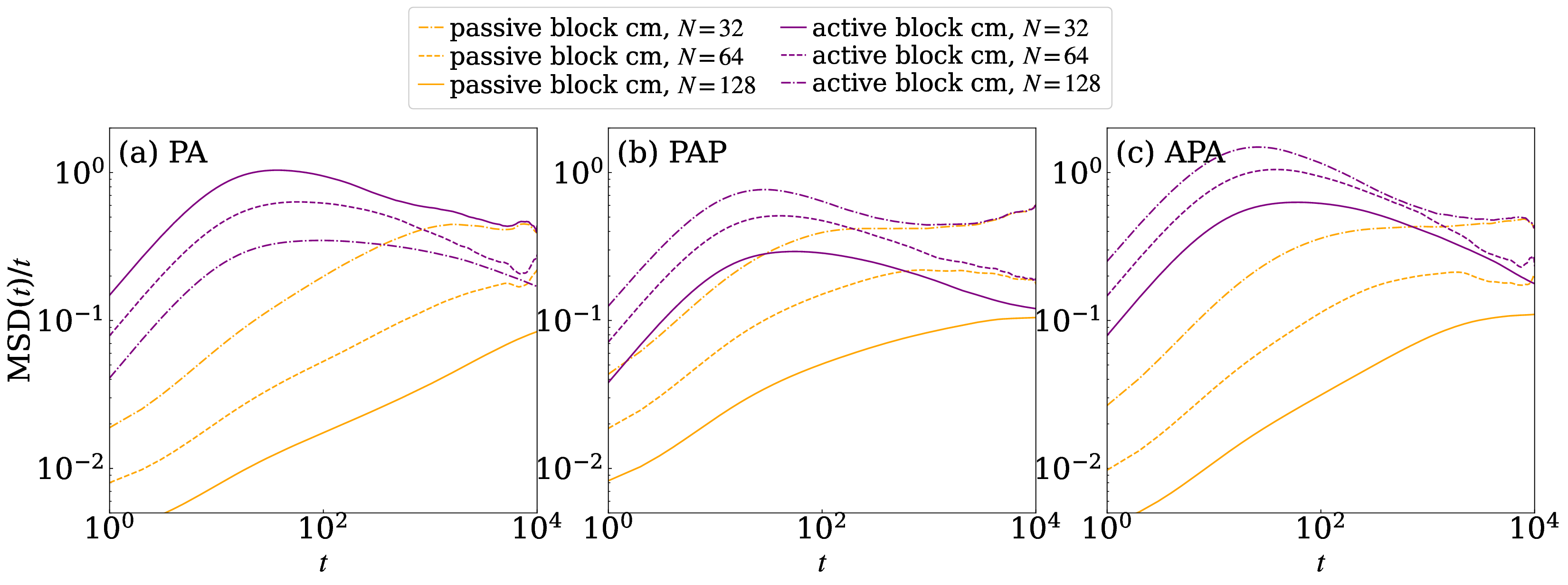}\\
\caption{\label{MSD_diffN} Chain-length dependence of ${\rm MSD}(t)/t$ for the cm of the passive and active blocks for different copolymers at ${\rm Pe=50}$, on a double-log scale.}
\end{figure*}
\par
The unusual conformational behavior naturally indulges us to probe the dynamics of the center of mass (cm) of the full polymer and the individual blocks. From their respective trajectories we calculate the corresponding mean square displacements 
\begin{equation}\label{MSD_formula}
{\rm MSD}(t) = \langle \left[\vec{r}_{\rm cm}(t)-\vec{r}_{\rm cm}(0)\right]^2\rangle, 
\end{equation}
where $\vec{r}_{\rm cm}$ is the position of the cm. Typically, MSD at short time captures a ballistic motion followed by normal diffusion with 
\begin{equation}\label{MSD_normal}
{\rm MSD}(t)=6D_{\rm eff}t,
\end{equation}
where $D_{\rm eff}$ is the effective diffusion constant. Unlike typical diffusion, anomalous diffusion is described by 
\begin{equation}\label{MSD_abnormal}
{\rm MSD}(t) = D_{\rm g}t^{\alpha},
\end{equation}
where $D_{\rm g}$ is the generalized diffusion constant  \cite{bouchaud1990,metzler2000,metzler2014}. The exponent $\alpha$ determines if it is a super-diffusion($\alpha >1$) or sub-diffusion($\alpha <1$). A typical diffusive behavior is obeyed by the data for FAP with $N=64$ in Fig.\ \ref{MSD}, where for the convenience of identification of different regimes we plot the time dependence of ${\rm MSD}/t$. The data show an initial linear increase followed by a plateau highlighting the diffusive regime. Similar behavior is also observed for the motion of the cm of the full polymer (red line) for all the copolymers, as presented in Figs.\ \ref{MSD}(a)-(c). In all cases, the length of the copolymers is chosen to be exactly twice than that of the FAP such that they contain the same number of active monomers. The plateau for all the copolymers correspond to the same value of $D_{\rm eff}$, which is significantly smaller than the corresponding $D_{\rm eff}$ for the FAP, indicating a strong influence of the passive blocks on the dynamics. 
\begin{figure*}[t!]
\centering
\includegraphics*[width=0.95\textwidth]{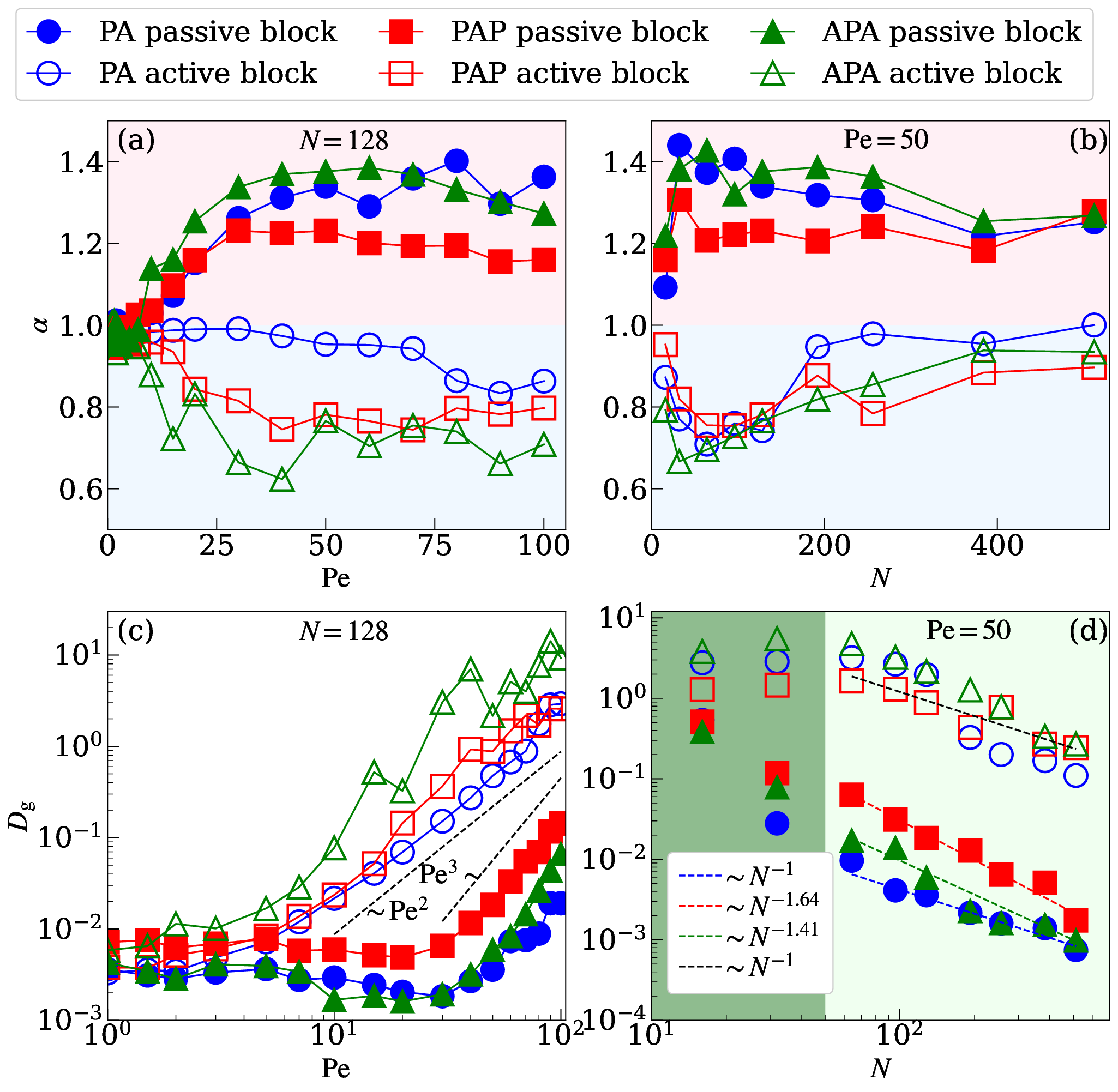}\\
\caption{\label{Diffusion_blocks} Dependence of the exponent $\alpha$ characterizing the anomalous diffusion on (a) the activity strength ${\rm Pe}$ and (b) chain length $N$. The shades there separates the sub-diffusive and super-diffusive behaviors. (c) and (d) illustrate the corresponding variation of the generalized diffusion constant $D_{\rm g}$. The dashed lines in (c) represent power-law behaviors as quoted next to them. The shades in (d) provides a guide to identify the power-law scaling regimes. The dashed lines in (d) represent respective fitted power-law scalings. The error bars are of the order of symbol size.}
\end{figure*}
\par
To disentangle the dynamics of the cm of the passive and active blocks we present their respective MSDs in Fig.\ \ref{MSD}. Although in the long-time limit ($t > 10^4$) MSD data of both blocks for all cases merge with the data of the full polymer, the preceding transient period show anomalous diffusion. There, the data for the passive block show a super-diffusive behavior with $\alpha \in [1.35,1.37]$, while the active block show sub-diffusion with $\alpha \approx 0.77$. Such anomalous diffusive behavior is a signature of intra-cellular transport phenomena \cite{caspi2000,Toli2004,bronstein2009,reverey2015,chen2015,lampo2017,song2018,molina2018}. In most cases this has been attributed to the crowded environment of the cell experienced by the probed bio-entity \cite{banks2005,jeon2011}. In the present case, however, this simply is a virtue of the tug-of-war between the passive and the active blocks. 
\par
In Fig.\ \ref{MSD_diffN} we present the chain-length dependence of the anomalous behavior of the ${\rm MSD}(t)$ for passive and active blocks. Clearly, data for all chain lengths show similar behavior as presented in Fig.\ \ref{MSD}. The data for the passive blocks for all the copolymers show pronounced super-diffusive behavior. The corresponding sub-diffusive behavior of the active blocks are comparatively less pronounced for PA, still deviating significantly from a normal diffusion. One also notices that for all the copolymers the data of the active blocks for the shortest chain length ($N=32$) merge with the data for corresponding the passive block at late time, when the full chain starts diffusing.. This merging occurs even later for longer chains. Importantly, both the passive and active blocks show a monotonic decrease of the amplitude of ${\rm MSD}(t)$ with increaseing chain length. This urges us to check the presence of any scaling of the generalized diffusion constant $D_{\rm g}$ as a function of $N$, which will be presented subsequently.
\par
To dig deep into the anomalous diffusion we analyze the power-law behavior of ${\rm MSD}(t)$ in the transient regime by calculating the exponent as 
\begin{equation}\label{alpha}
\alpha=\left \langle \frac{d \ln {\rm MSD}(t)}{d \ln t}\right \rangle,
\end{equation}
where the $\langle \dots \rangle$ indicates an average over different times within the transient regime as well as different trajectories. In Fig.\ \ref{Diffusion_blocks}(a) we present the variation of $\alpha$ with ${\rm Pe}$, for a fixed $N$ for both the blocks of different copolymers. It shows how starting from $\alpha=1$ the super- and sub-diffusion emerge, respectively, for the passive and active blocks as ${\rm Pe}$ increases. In the large ${\rm Pe}$ limit, $\alpha$ for the super-diffusion settles at a slightly smaller value for PAP compared to the other copolymers. The sub-diffusive behavior of the active blocks do not show the same trend, and roughly settles around $0.8$ for all of them. On the other hand, one can notice from Fig.\ \ref{Diffusion_blocks}(b) that for long $N$ the super-diffusive $\alpha$ is almost the same for all the copolymers, as also is the case for $\alpha$ for the sub-diffusion of the active blocks. 

\par
Having established the evidence of anomalous diffusion in the transient regime, we calculate the corresponding generalized diffusion constant as 
\begin{equation}
 D_{\rm g}=\left \langle \exp\left[\frac{\ln t \ln {\rm MSD}(t^{\prime}) -\ln t^\prime\ln {\rm MSD}(t)}{\ln t-\ln t^{\prime}}\right] \right \rangle,
\end{equation}
where the times $t$ and $t^\prime$ are within the transient regime, and $\langle \dots \rangle$ denotes averaging over different $(t,t^{\prime})$ pairs and independent trajectories. The corresponding plots for $D_{\rm g}$ as a function of ${\rm Pe}$ is shown in Fig.\ \ref{Diffusion_blocks}(c) for all the copolymers. For smaller ${\rm Pe}$, $D_{\rm g}$ remains almost invariant for all cases. At around  ${\rm Pe}\approx 10$, the active blocks show a transition to a ${\rm Pe}$-dependent behavior. The passive blocks show a similar transition at larger ${\rm Pe} \approx 25$. Even though the active blocks show the usual $D_{\rm g} \sim {\rm Pe}^2$ dependence \cite{majumder2024}, the behavior of the passive block is even more enhanced with $D_{\rm g} \sim {\rm Pe}^3$. The corresponding scaling behaviors with respect to the chain length $N$ are  shown in Fig.\ \ref{Diffusion_blocks}(d). The data for the passive blocks show a lot of diversity with only the behavior for PA being roughly consistent with a Rouse-like scaling $D_{\rm g}\sim N^{-1}$ \cite{rouse1953}. Both PAP and APA show anomalous behavior with even slower dynamics. In contrast, the data for the active blocks of all the copolymers are more or less consistent with the Rouse-like scaling \cite{majumder2024}. Of course, a proper theoretical treatment is required to confirm the validity of the observed scaling. Nevertheless, it can be inferred that the dynamics in the transient regime is rich, and is highlighted by the anomalous diffusion and related scaling.
\begin{figure}[t!]
\centering
\includegraphics*[width=0.48\textwidth]{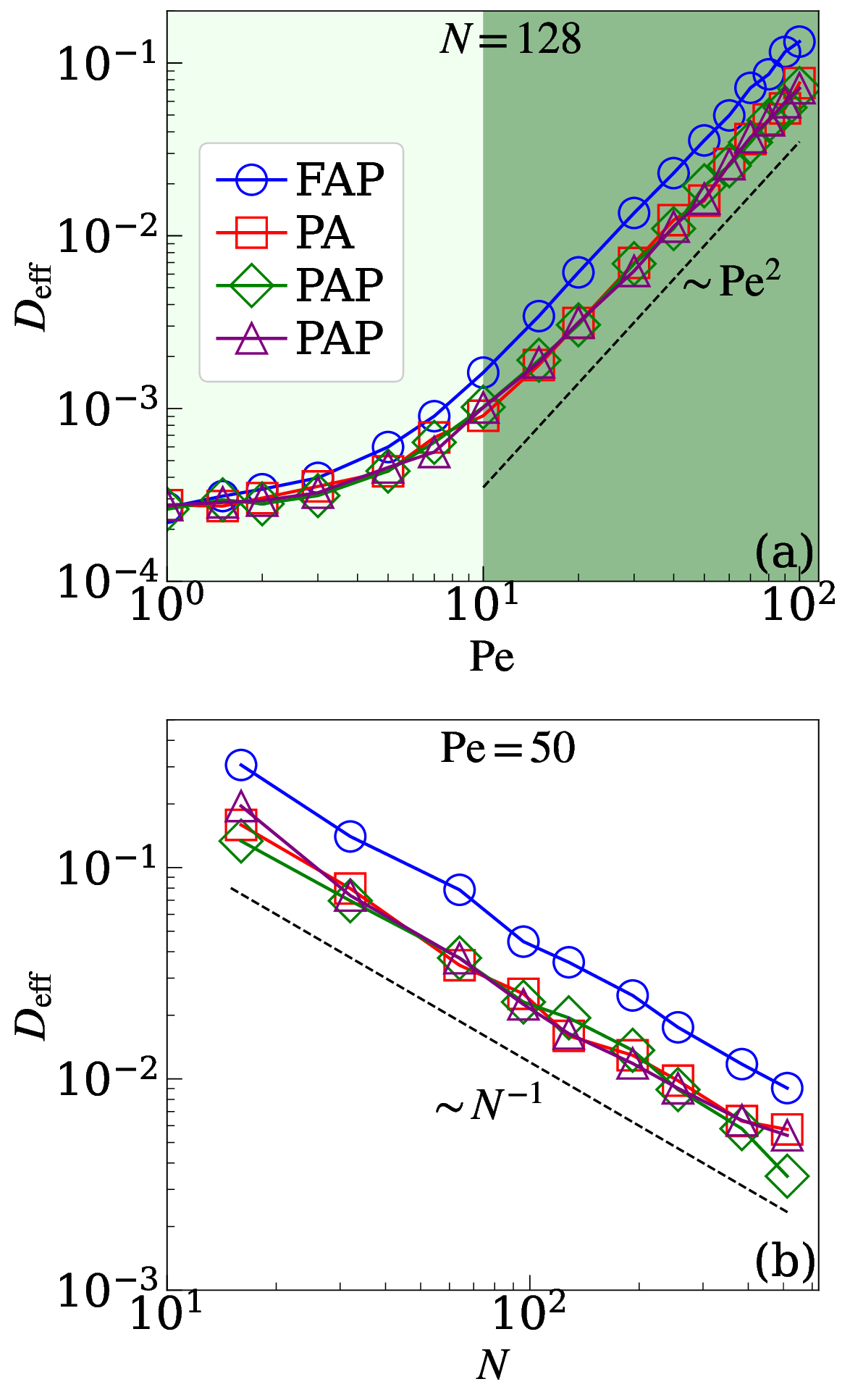}
\caption{\label{Diffusion_CM} Dependence of the effective diffusion constant of the cm of the full polymer $D_{\rm eff}$ for different types of copolymer as a function of (a) ${\rm Pe}$ and (b) $N$. In (a) the data is for a fixed $N=128$ and (b) is for a fixed ${\rm Pe}=50$. For a comparison corresponding data for FAP are also shown. The dashed line in (a) correspond to a quadratic behavior whereas the dashed line in (b) represents the Rouse-like scaling $D_{\rm eff} \sim N^{-1}$. Different shades in (a) provide a guide to identify the power-law regime.}
\end{figure}
\par
Finally, to check if the anomalous transient dynamics of the individual blocks have any effect on the long-time diffusion of the full polymer, we calculate the effective diffusion constant of the cm of polymer as \begin{equation}
D_{\rm eff}=\left \langle \frac{1}{6}\lim_{t \rightarrow \infty}\frac{d}{dt}{\rm MSD}(t)\right \rangle.                                                            \end{equation}
The variation of $D_{\rm eff}$ with ${\rm Pe}$ is shown in Fig.\ \ref{Diffusion_CM}(a), which also includes the data for a FAP. All the copolymers show behavior similar to that of a FAP. For ${\rm Pe} < 10$, $D_{\rm eff}$ varies marginally, and beyond that it starts showing a quadratic dependence \cite{bianco2018,majumder2024}. The crossover occurs almost at the same value of ${\rm Pe}$ where $D_{\rm g}$ of the active blocks show a transition to a quadratic behavior as shown in Fig.\ \ref{Diffusion_blocks}(c). The corresponding scaling with respect to the chain length $N$ is presented in Fig.\ \ref{Diffusion_CM}(b) showing again a behavior proportional to the data for a FAP. In other words, the enhanced diffusion constant obeys a Rouse-like scaling $D_{\rm eff}\sim N^{-1}$ for all the copolymers \cite{majumder2024}.
\section{Conclusion}\label{conclusion}
To summarize, using numerical simulations we have presented results for the conformation and dynamics of linear active block copolymers where blocks of passive and active monomers reside along its contour. Our results show that depending on the respective positions of passive and active blocks one can tune the amount of swelling achieved by the polymer due to the exerted activities. For example, the copolymer APA where two active blocks are present at the two ends with a passive block in between, exhibits more swelling than a fully active polymer of same length. This is attributed to the pulling of the passive block by the two connected active blocks leading to an expansion of the passive block. The behavior of the other copolymers can also be argued on the basis of the tug-of-war between the passive and active blocks. This fact possibly can explain similar peculiarities in conformations and dynamics, and thereby the functionality of many bio-polymers. Even though the long-time dynamics of the cm of the copolymers exhibit the usual diffusive Rouse-like scaling, the long-lived transient regime  of the passive and active blocks, respectively, show anomalous super- and sub-diffusion. The corresponding generalized diffusion constant also exhibits non-trivial scaling behaviors with the chain length. Considering the recent progress in developing synthetic active polymers \cite{biswas2017}, the results presented here should indulge design of new polymeric materials with tailored  static and dynamic properties depending on the relative position of the passive and active blocks. 
\par
In future, it would be interesting to explore other patterns of relative arrangements of the passive and active blocks. Sequences mimicking real chromosomal patterns would reveal the relevance of the anomalous features observed here with such dynamics pertinent to motion of cell organelles within the cell nucleus. From a technical point of view it would also be intriguing to include the role of inertia \cite{lowen2020}, which for particle system triggers orientational ordering \cite{paul2024}.

\begin{acknowledgments}
The work was funded by the Science and Engineering Research Board (SERB), Govt.\ of India for a Ramanujan Fellowship (file no.\ RJF/2021/000044). S.P.\ acknowledges University of Delhi for providing financial assistance through the Faculty Research Programme Grant-IOE (ref.\ no.\ /IOE/2024-25/12/FRP).
\end{acknowledgments}

\end{document}